\documentclass[pdflatex,sn-mathphys-num]{sn-jnl}

\usepackage{graphicx}%
\usepackage{multirow}%
\usepackage{amsmath,amssymb,amsfonts}%
\usepackage{amsthm}%
\usepackage{mathrsfs}%
\usepackage[title]{appendix}%
\usepackage{xcolor}%
\usepackage{textcomp}%
\usepackage{manyfoot}%
\usepackage{booktabs}%
\usepackage{algorithm}%
\usepackage{algorithmicx}%
\usepackage{algpseudocode}%
\usepackage{listings}%

\theoremstyle{thmstyleone}%
\newtheorem{theorem}{Theorem}
\newtheorem{proposition}[theorem]{Proposition}%

\theoremstyle{thmstyletwo}%
\newtheorem{example}{Example}%
\newtheorem{remark}{Remark}%

\theoremstyle{thmstylethree}%
\newtheorem{definition}{Definition}%

\def\be{\begin{eqnarray}}
\def\ee{\end{eqnarray}}

\newcommand{\ZGC}{{\cal Z}}
\newcommand{\besK}{\mathrm{K}}
\newcommand{\diff}{\mathrm{d}}

\begin{document}

\title[BQ correlation at finite density]{$S$-matrix calculation of $BQ$ correlation at finite baryon density}


\author[1]{\fnm{Vojt\v{e}ch} \sur{Hon\v{e}k}}

\author[2]{\fnm{Pok Man} \sur{Lo}}

\author[1,3]{\fnm{Boris} \sur{Tom\'a\v{s}ik}}

\affil[1]{\orgdiv{Fakulta jadern\'a a fyzik\'aln\v{e} in\v{z}en\'yrsk\'a}, \orgname{\v{C}esk\'e vysok\'e u\v{c}en\'i technick\'e v Praze}, \orgaddress{\street{B\v{r}ehov\'a 7}, \city{Praha 1}, \postcode{11519}, \country{Czech Republic}}}

\affil[2]{\orgdiv{Institute of Theoretical Physics}, \orgname{University of Wroc{\l}aw}, \orgaddress{\street{plac Maksa Borna 9}, \city{Wroc{\l}aw}, \postcode{PL-50204}, \country{Poland}}}

\affil[3]{\orgdiv{Fakulta pr\'irodn\'ych vied}, \orgname{Univerzita Mateja Bela}, \orgaddress{\street{Tajovsk\'eho 40}, \city{Bansk\'a Bystrica}, \postcode{97401},  \country{Slovakia}}}


\abstract{We calculate the baryon number--electric charge susceptibility at non-vanishing baryo-chemical potential within the model of hadron gas where pion-nucleon interaction is accounted for by the $S$-matrix formalism.
The susceptibility is largely increased when the chemical potential grows within a phenomenologically relevant interval. 
The results are then evaluated along the chemical freeze-out line. 
We also calculate the evolution of the susceptibility in a cooling fireball by making use of the Partial Chemical Equilibrium model. 
}




\maketitle

\section{Introduction}
\label{s:intro}

A gas of interacting hadrons provides good description of the thermodynamics of hot strongly interacting matter below the deconfinement  transition temperature.
The model has recently been improved by including the interactions with the help of $S$-matrix formalism \cite{Dashen:1969ep,Weinhold:1997ig,Lo:2017ldt}. 
This step has led to better agreement with the lattice QCD calculations \cite{Lo:2017lym}. 
On the phenomenological side, it removed an earlier tension between proton yields from Pb+Pb collisions at $\sqrt{s_{NN}} = 2.76$ TeV and the calculations using the Hadron Resonance Gas (HRG) model \cite{Stachel:2013zma,Andronic:2018qqt}. 

It has also been shown that the correlation between electric charge and baryon number quantified by the susceptibility $\chi_{BQ}$ and calculated within the $S$-matrix formalism for the $\pi$N scattering becomes better aligned with the lattice QCD results \cite{Lo:2017lym}. 
The comparison has been done at vanishing net-baryon density, where lattice QCD calculations are directly applicable.

On the other hand, fluctuations and correlations of conserved quantum charges at finite baryon density are phenomenologically interesting as they are sensitive to the vicinity of a critical point in the phase diagram. 
Therefore, in this paper we extend the calculation of $\chi_{BQ}$ to non-vanishing baryo-chemical potential and discuss also its possible evolution during expansion and cooling of the hot matter in ultrarelativistic heavy-ion collisions.


\section{Model and methods}
\label{s:model}

For clarity, we start by splitting the partition function into two parts: free ground state zero-width hadrons and contribution from interactions between them. 
The two parts will be denoted 
\begin{equation}
\ln \ZGC = \ln \ZGC_0 + \ln \ZGC_{\mathrm{int}}\, .
\end{equation}
The free part of the partition function includes all hadrons from the meson octet as well as the baryon octet and decuplet that are stable with respect to strong interactions. 
Both baryons and antibaryons are included. 
It is then
\begin{equation}
\label{e:ZGC0}
\ln \ZGC_0 = \frac{VT}{2\pi^2} \sum_i g_i m_i^2  \sum_{n=1}^{\infty} \frac{(\mp 1)^{n-1}}{n^2} e^{n\mu_i/T} \besK_2\left ( \frac{n m_i}{T} \right )\, .
\end{equation}
The first sum runs over all included species and the second comes from the expansion of quantum-statistical distribution (FD or BE).
They sum up contributions from species with mass $m_i$ and spin degeneracy $g_i$, which stem from a system with volume $V$ at temperature $T$.
The contributions are weighted by the modified Bessel functions $\besK_2$. 
In chemical equilibrium, the chemical potential of species $i$ will be determined as
\begin{equation}
\mu_i = B_i \mu_B + S_i \mu_S + Q_i \mu_Q\, ,
\end{equation}
where $B_i$, $S_i$, $Q_i$ are the baryon number, strangeness, and electric charge of species $i$, and $\mu_B$, $\mu_S$, $\mu_Q$ are chemical potentials attached to these quantum numbers, respectively.
Later on, when we describe the evolution of a cooling fireball, this coupling of $\mu_i$ to chemical potentials of conserved quantum numbers will be relaxed and each ground state hadron species will acquire its own chemical potential.

The interaction part of the partition function is generally  given as
\begin{multline}
\label{e:ZGCi}
\ln \ZGC_{\mathrm{int}} = \frac{VT}{2\pi^2}
\sum_{B,S,I}
\left | \langle I^{(1)}, I^{(1)}_z, I^{(2)}, I_z^{(2)} | I , I_z \rangle\right |^2 
\\ \times
\sum_J g_J \sum_{n=1}^\infty \frac{(\mp 1)^{n-1}}{n^2} \exp\left ( \frac{n\mu_{\mathrm{int}}}{T}\right )
\int \diff M\, M^2\, \frac{1}{\pi} \frac{\diff\delta_{I,J}(M)}{\diff M}  
\besK_2\left ( \frac{n M}{T} \right )\, .
\end{multline}
The first sum runs over states with baryon number $B=-1,0,1$, strangeness $S=-3,-2,\dots,3$, and all isospin states. 
Each channel is multiplied with the Clebsch-Gordan coefficient for the combination of isospins. 
Under isospin symmetry, all coefficients would add up to 1, but the coefficients must be explicitly included if particle numbers are to be calculated. 
Next, the sum over $J$ adds up partial waves with different angular momenta, each multiplied with the appropriate spin degeneracy $g_J$. 
As in eq.~(\ref{e:ZGC0}), the sum over $n$ comes from expanding the FD or BE distribution. 
The chemical potential is given by the sum of the chemical potentials of the interacting species.
Finally, the integral over $M$ runs over the derivative of the phase shifts $\delta_{I,J}$ for each isospin-spin channel, weighted by the statistical distribution which results in the modified Bessel function $\besK_2$. 

If the phase shift is assumed to be a stepwise function, then ${\diff\delta_{I,J}(M)}/{\diff M}$ yields $\delta$-function and the formula (\ref{e:ZGCi}) reduces to a sum over zero-width resonances~\footnote{
A consistent treatment of the thermal contribution of a finite-width state requires both the full spectral function and the non-resonant background. Restricting to the former, as done in most previous works under the Breit–Wigner approximation, introduces inconsistencies and model dependence (see Ref.~\cite{Lo:2019who}).}
which is formally analogous to eq.~(\ref{e:ZGC0})
\begin{multline}
\ln \ZGC_{\mathrm{int}} \to \ln\ZGC_R \\
= \sum_{B,S,I} 
\left | \langle I^{(1)}, I^{(1)}_z, I^{(2)}, I_z^{(2)} | I , I_z \rangle\right |^2 
\sum_J g_J \sum_{n=1}^\infty \frac{(\mp 1)^{n-1}}{n^2} \exp\left ( \frac{n\mu_{\mathrm{int}}}{T}\right )
\\ \times
\sum_{R'(I,J)} 
\int \diff M\, M^2\, \delta(M-m_R)  
\besK_2\left ( \frac{n M}{T} \right )\\
= \frac{VT}{2\pi^2} \sum_R g_R m_R^2  \sum_{n=1}^{\infty} \frac{(\mp 1)^{n-1}}{n^2} e^{n\mu_R/T} \besK_2\left ( \frac{n m_R}{T} \right )\, .
\end{multline}
Note that the sum $\sum_{R'(I,J)}$ counts all resonance states with given isospin and spin whereas $\sum_R$ includes all resonances with all spins and isospins. 
The chemical potential of the interaction channel becomes that of the resonance. 
This description will be used for all mesonic and baryonic resonances, except those arising from pion-nucleon scattering, i.e. unflavored baryonic channels.

The latter will be implemented with explicit phase shifts from the GWU/SAID partial wave analysis \cite{Workman:2012hx}, as was done in \cite{Lo:2017lym}.
By explicitly writing the isospins in the CG coefficients, the contribution to the partition function will be
\begin{multline}
\label{e:ZGCps}
\ln \ZGC_{\mathrm{int}} \to \ln \ZGC_i \\
= \frac{VT}{2\pi^2}
\sum_{\substack{I^\pi_z,I^N_z,I \\ B=-1,1}} 
\left | \langle 1, I^{\pi}_z, \textstyle{\frac{1}{2}}, I_z^{N} | I , I_z \rangle\right |^2 
\sum_J g_J \sum_{n=1}^\infty \frac{(- 1)^{n-1}}{n^2} \exp\left ( \frac{n\mu_{\mathrm{int}}}{T}\right )
\\ \times
\int \diff M\, M^2\, \frac{1}{\pi} \frac{\diff\delta_{I,J}(M)}{\diff M}  
\besK_2\left ( \frac{n M}{T} \right )\, .
\end{multline}
For the $\pi N$ interaction, 
the first sum explicitly sums over total isospins $I=1/2$ and $I=3/2$.
In other words, the $N^*$ and $\Delta$ resonances are included via the phase shifts. 
In such a case, the chemical potential in the interaction term simply becomes
\begin{equation}
\mu_{\mathrm{int}} = \mu_{\pi,I^\pi_z} + \mu_{N,I^N_z}\, .
\end{equation}
In chemical equilibrium, by using the Gell-Mann--Nishijima formula, this becomes
\begin{equation}
\mu_{\mathrm{int}} = \left ( I_\pi^z + I_N^z + \frac{B}{2} \right ) \mu_Q + B \mu_B\,  .
\label{e:mueq}
\end{equation}
Furthermore, interaction between $\eta$ and $N$ is also included via the phase-shifts, as was done in \cite{Lo:2017lym}.

The last interaction to include are three-particle interactions of $\pi\pi N$. 
Such final states appear in decays of $N^*$ and $\Delta$ resonances, but are not included in the phase-shifts contributions. 
Thus, they have to be accounted for in the partition function differently. 
Here, we have to resort to an effective treatment to include their  contribution. 

The $\pi\pi N$ interactions can be partially accounted for by modifying the properties of the intermediate $\Delta(1232)$ resonance~\cite{Lo:2022nhk}. To avoid double counting, the corresponding $\Delta$-exchange contribution already included in $\pi N \to \pi N$ via the $P_{33}$ phase shift must be subtracted. In the following, we consider three implementations of this class of models. We perform calculations with all of them and use the spread of the results as an estimate of the associated systematic uncertainty.

\begin{description}
    \item[$\Delta(1232)$ model:]
    The $\pi\pi N$ interactions are accounted for by a modification of the $\Delta(1232)$ resonance.
    The modified resonance is shifted in mass and added to the $\ln \ZGC_R$ contribution.
    At the same time, $\Delta(1232)$ with its original mass is subtracted from $\ln \ZGC_R$ in order to prevent the double-counting, as it has been also included in the phase-shifts part. 
    The shifted mass is 1.1799~GeV and is chosen so that $\chi_{BQ}$ at $T=155$~MeV and $\mu_B=0$ reproduces the lattice value~\cite{Bollweg:2021vqf}. The same is done for the $B=-1$ resonance. 
    \item[Shifted $N^*$ and $\Delta$ 1:] 
    Three-body interactions are included as modified $N^*$ and $\Delta$ resonances. 
    Their degeneracies are reduced to the parts which correspond to only the three-body branching ratios. 
    The masses are all multiplied by common factors: one for $N^*$'s and other for $\Delta$'s. 
    By fitting to lattice QCD results on $\chi_{BQ}$~\cite{Bollweg:2021vqf} at $T=135$~MeV and 155~MeV these factors are set to 0.966 for $N^*$'s and 1.20 for $\Delta$'s.
    \item[Shifted $N^*$ and $\Delta$ 2:] 
    The model is constructed in the same way as the previous one, but lattice QCD results are fitted in a larger intesval of temperatures, from $T=135$~MeV to 160~MeV. 
    The resulting factors are 0.952 for $N^*$'s and 2.16 for $\Delta$'s.
\end{description}

Hence, after acknowledging technically different treatment of interactions in $\pi N$ and $\eta N$ channel from the remaining ones, and effectively including the $\pi\pi N$ interactions, the full partition function is written 
\begin{equation}
    \ln \ZGC = \ln \ZGC_0 + \ln \ZGC_R + \ln \ZGC_i + \ln \ZGC_3\, .
\label{e:Zcont}    
\end{equation}
The last term, even though it is technically identical to $\ln \ZGC_R$, is pointed out here, since we will be interested  in its particular contribution. 
Since we are mostly interested in the actual contributions of the last two terms, we will denote the results obtained from the first two terms as `baseline'. 

Thermodynamic quantities are then obtained by taking the appropriate derivatives of the partition function. 
We will need the entropy density
\begin{equation}
s = \frac{1}{T} \left ( p  + \varepsilon - \sum_i n_i \mu_i \right )
\label{e:sGD}
\end{equation}
where the pressure, energy density, and number density of species $i$ are calculated as 
\begin{eqnarray}
p & = & \frac{T}{V} \ln\ZGC\, \\
\varepsilon & = & - \frac{1}{V} \frac{\partial \ln \ZGC}{\partial \beta}\, \\
n_i & = & \frac{1}{V}\frac{\partial \ln \ZGC}{\partial \alpha_i}\, .
\label{e:ndens}
\end{eqnarray}
Here, $\alpha_i = \mu_i/T$, and $\beta = 1/T$. 
The $BQ$ susceptibility
\begin{equation}
\chi_{BQ} = \frac{T}{V} \frac{\partial^2 \ln \ZGC}{\partial \mu_B \partial \mu_Q}\, ,
\end{equation}
is then expressed as 
\begin{multline}
\chi_{BQ} =  \frac{1}{2\pi^2} \sum_{i,R,3} B_i Q_i g_i m_i^2  \sum_{n=1}^{\infty} (\mp 1)^{n-1} e^{n\mu_i/T} \besK_2\left ( \frac{n m_i}{T} \right )
\\
+ \frac{1}{2\pi^2} \sum_{\substack{I^{\alpha}_z,I^N_z,I_z \\ B=-1,1}} 
B Q_{I_z}
\left | \langle I^\alpha, I^{\alpha}_z, {\textstyle{\frac{1}{2}}}, I_z^N | I , I_z \rangle\right |^2 
\\ \times
\sum_J g_J \sum_{n=1}^\infty (- 1)^{n-1} \exp\left ( \frac{n\mu_{I_z}}{T}\right )
\\ \times
\int \diff M\, M^2\, \frac{1}{\pi} \frac{\diff\delta_{I,J}(M)}{\diff M}  
\besK_2\left ( \frac{n M}{T} \right )\, .
\end{multline}
Here, the first sum counts the ground state hadrons and zero width resonances, while the second sum accounts for interactions included via phase shifts. 
The symbol $\alpha$ represents $\pi$ or $\eta$ meson, and $I^\alpha$ is its total isospin.
The last formula is written in a way that remains valid even off the chemical equilibrium, where eq.(\ref{e:mueq}) does not hold. 

\begin{figure}
\begin{center}
\includegraphics[width=0.8\textwidth]{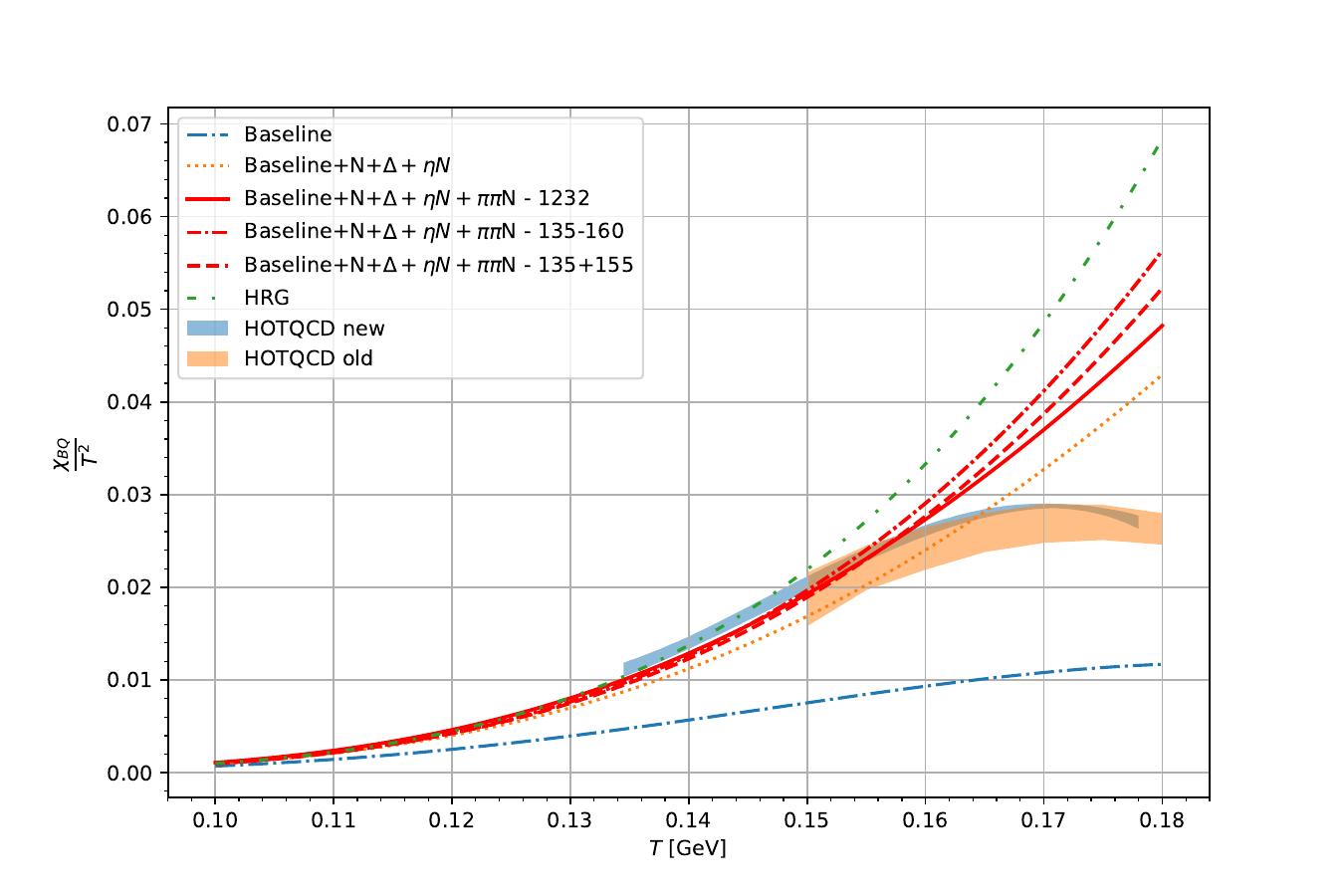}
\end{center}
\caption{The scaled susceptibility $\chi_{BQ}/T^2$ as function of temperature for $\mu_B =0$.
Blue dash-dotted line denoted 'Baseline' refers to the result from the first two contributions in Eq.~(\ref{e:Zcont}). Orange dotted line: contributions via phase-shifts added. Red lines refer to the three models that represent $\pi\pi N$ interactions: $\Delta(1232)$ model (solid), Shifted $N^*$ and $\Delta$ 1 (dash-dotted), Shifted $N^*$ and $\Delta$ 2 (dashed). The Hadron Resonance Gas result is shown by blue short-dashed-dotted curve. Lattice QCD data from newest publication \cite{Bollweg:2021vqf} (blue, HOTQCD new) and earlier publication \cite{HotQCD:2012fhj} (orange, HOTQCD old) are also shown.
}
\label{f:chiB0}
\end{figure}
In Fig.~\ref{f:chiB0} we illustrate how the models describe  $\chi_{BQ}$ at vanishing baryo-chemical potential. 
Note that in comparison to \cite{Lo:2017lym} we use a new issue of lattice QCD results~\cite{Bollweg:2021vqf} to which we tune our three-body contribution. 
For illustration, both old \cite{HotQCD:2012fhj} and new sets of lattice data are included in the figure. 
Later on, we will present as our primary results those obtained with the $\Delta(1232)$ model.
The different predictions from the three models of $\pi\pi N$ interactions  will be treated as representing the systematic uncertainty of our calculations. 


\section{Results for the $BQ$ correlation}
\label{s:results}

\begin{figure}
\begin{center}
\includegraphics[width=0.8\textwidth]{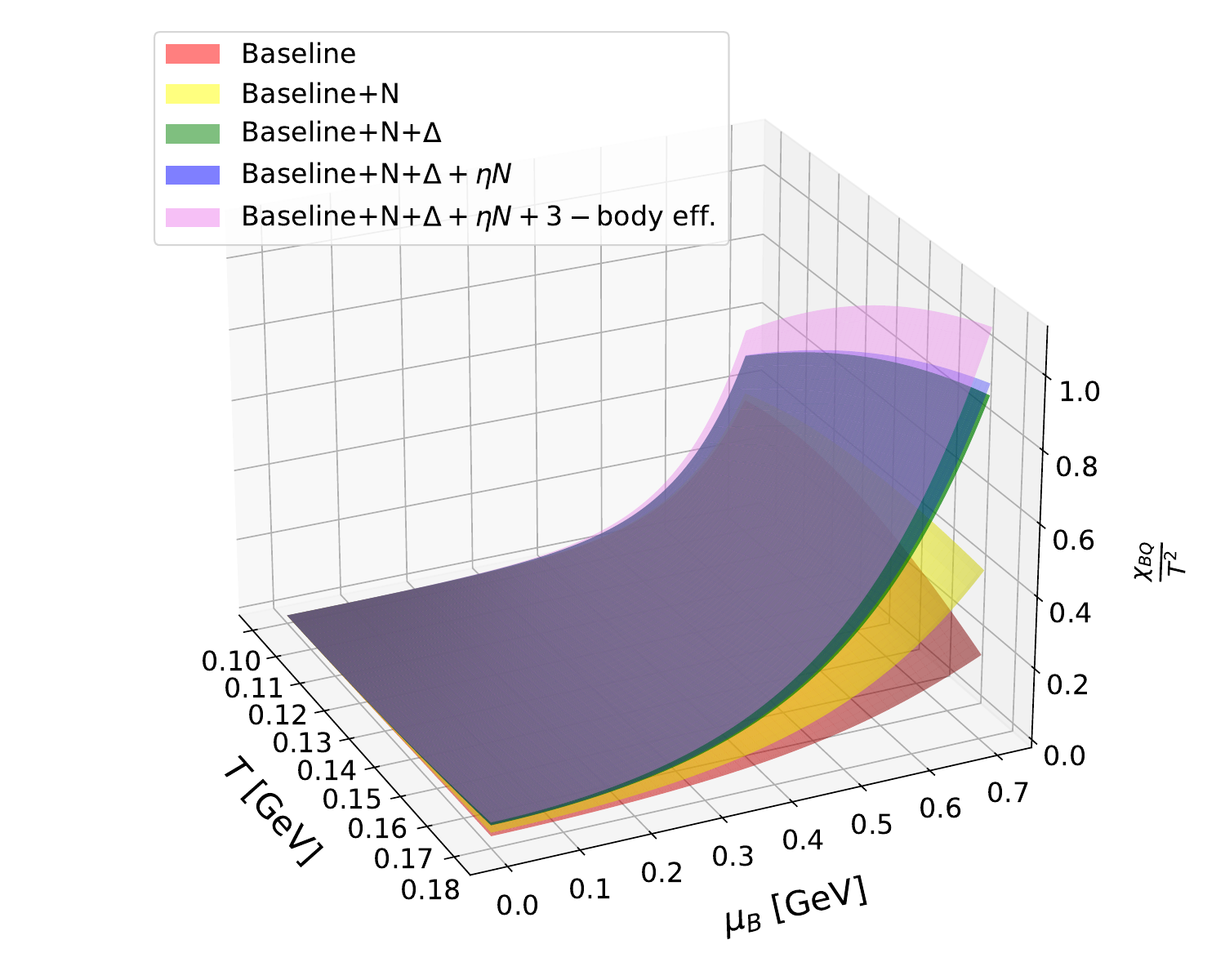}
\end{center}
\caption{The scaled susceptibility $\chi_{BQ}/T^2$ as function of temperature and baryo-chemical potential. 
Red (lowest surface, `baseline'): no contributions from the phase shift part of the partition function. Yellow (`baseline + N'): channels with $I=1/2$ added in the phase shift interactions. Green (`baseline + N + $\Delta$'): also channels with $I=3/2$ added in the phase shift interactions. Blue (`baseline + N + $\Delta$ + $\eta N$'): also $\eta N$ channels added in the phase shift interactions. Purple (uppermost surface): complete calculation including also $\pi\pi N$ interactions.}
\label{f:chiB}
\end{figure}

The $BQ$ susceptibility has been calculated for non-va\-ni\-shing baryo-chemical potential, thereby extending the results of \cite{Lo:2017lym}.
The results are shown in Fig.~\ref{f:chiB}.
%
%
Note that the dimensionless quantity $\chi_{BQ}/T^2$ is plotted. 
The figure shows how the inclusion of interactions via phase shifts in $I=1/2$ and $I=3/2$ channels contributes to the correlation of baryon number and electric charge. 
At high chemical potential, the non-interacting part of the partition function shows a decreasing dependence of $\chi_{BQ}/T^2$ on  temperature, but adding the interaction makes that dependence increasing. 

Nevertheless, phenomenologically more relevant case would assume strangeness neutrality throughout the whole investigated medium. 
Introducing baryo-chemical potential breaks strangeness neutrality because it enhances ba\-ry\-ons with $S<0$ over antibaryons with $S>0$. 
Strangeness neutrality is then restored by introducing non-zero stran\-ge\-ness chemical potential which is implicitly determined for every pair of $T$ and $\mu_B$ via the requirement for vanishing net strangeness density
\begin{equation}
0 = n_S = \sum_i S_i\, n_i(T,\mu_B,\mu_S)\, ,
\end{equation}
where the sum goes over all involved species and $S_i$ is the strangeness of species $i$ and $n_i$ is calculated via eq.~(\ref{e:ndens}). 

\begin{figure}
\begin{center}
\includegraphics[width=0.8\textwidth]{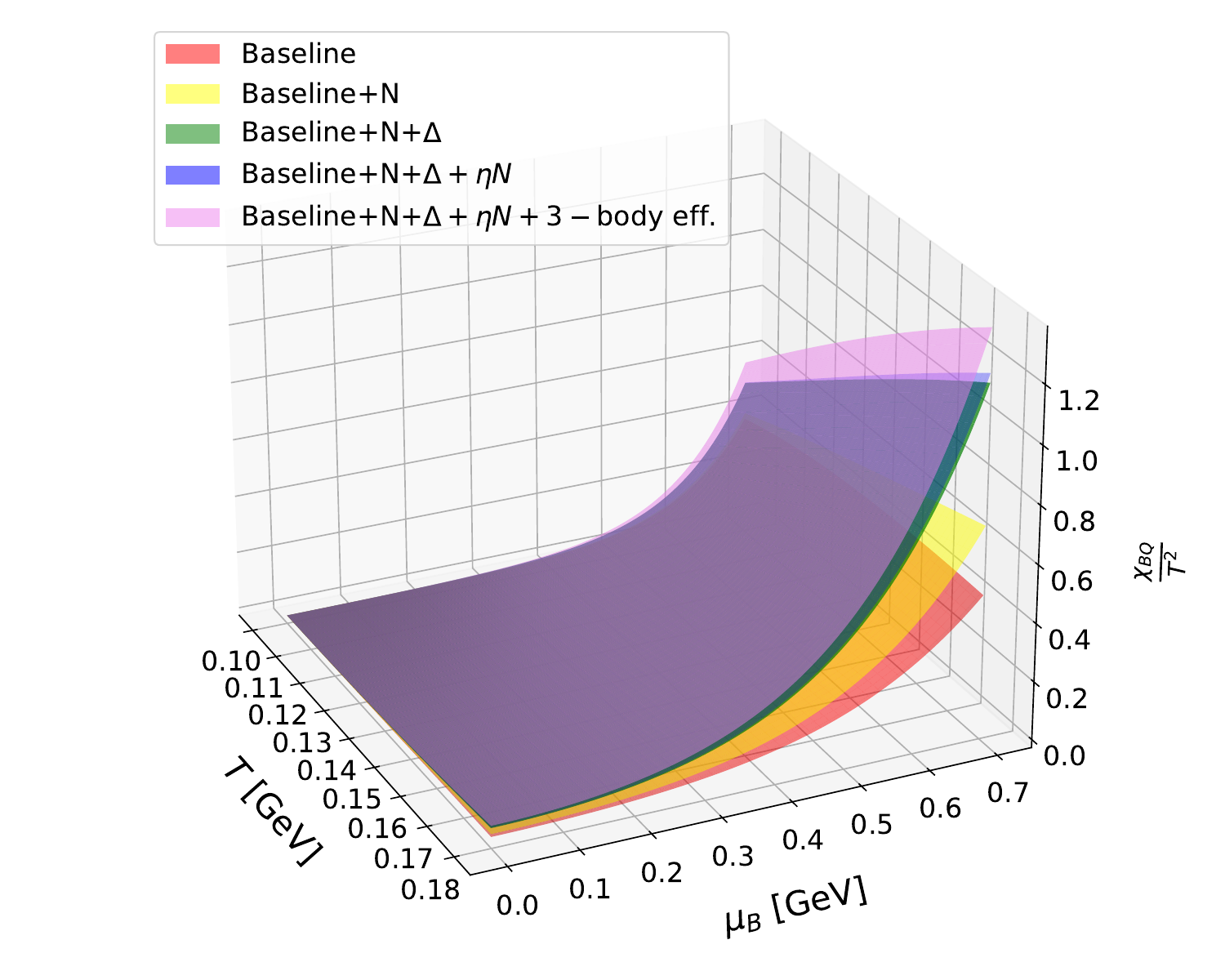}
\end{center}
\caption{Same as Fig.~\ref{f:chiB}, but calculated for matter with vanishing net strangeness.}
\label{f:chiS0}
\end{figure}

The $BQ$ susceptibility for strangeness-neutral matter is plotted in Fig.~\ref{f:chiS0}. 
%
%
In general, the correlation between $B$ and $Q$ enhances and the dependence of $\chi_{BQ}/T^2$ from the non-interacting contributions on $T$ does not show the dramatic decrease at high $\mu_B$.

\begin{figure}
\begin{center}
\includegraphics[width=0.65\textwidth]{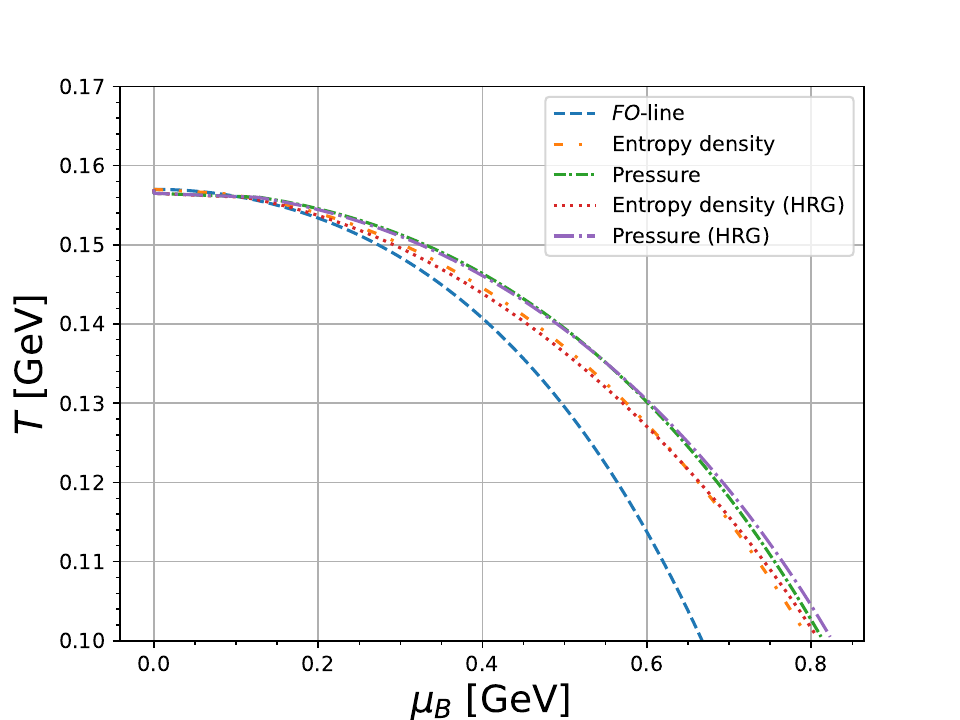}
\end{center}
\caption{Chemical freeze-out (FO) line as parametrised by eq.~(\ref{e:tfo}), lines of constant entropy density, and lines of constant pressure, calculated with interactions via $S$-matrix and within the HRG model.}
\label{f:fo}
\end{figure}

We are now ready to introduce the combinations  of temperature and chemical potentials that are indicated by fits to hadron abundances throughout different collision energies, as measured in collisions of heavy atomic nuclei, Pb+Pb or Au+Au. 
The chemical freeze-out temperature as a function of baryo-chemical potential can be parametrised as
\begin{equation}
T(\mu_B) = a - b\mu_B^2 - c\mu_B^4\, ,
\label{e:tfo}
\end{equation}
with $a=0.157$~GeV, $b = 0.087$~GeV$^{-1}$, and $c = 0.0092$~GeV$^{-3}$ \cite{Vovchenko:2015idt}.
In Fig.~\ref{f:fo} we compare this parametrisation with the lines that denote constant pressure or constant entropy density. 
%
%
These lines are calculated for the model introduced in the previous section and also for a Hadron Resonance Gas model where the $S$-matrix interaction partition function is replaced by corresponding resonances with vanishing widths. 
All curves start at the same temperature at $\mu_B = 0$. 
Generally, lines of constant pressure give slightly higher temperature for the same chemical potential than lines of constant entropy density. 
The two compared models yield very similar temperatures, though. 
The FO line decreases considerably faster as a function of $\mu_B$ than lines of constant pressure or entropy density. 

Note that the parametrisation (\ref{e:tfo}) was extracted from fits with the Hadron Resonance Gas model which does not treat the interactions through the phase shifts. 
Hence, its use introduces slight inconsistency in our treatment. 
Nevertheless, we do not expect a major change in $T$ and $\mu_B$ for a model with phase shifts, since also for the curves of constant entropy and pressure the two models also give very similar results. 
Moreover, an analysis of hadron abundances with our model goes far beyond the scope of this work. 

Now, the $BQ$-susceptibility can be calculated as a function of temperature along the lines shown in Fig.~\ref{f:fo}. 
%
\begin{figure}
\begin{center}
\includegraphics[width=0.48\textwidth]{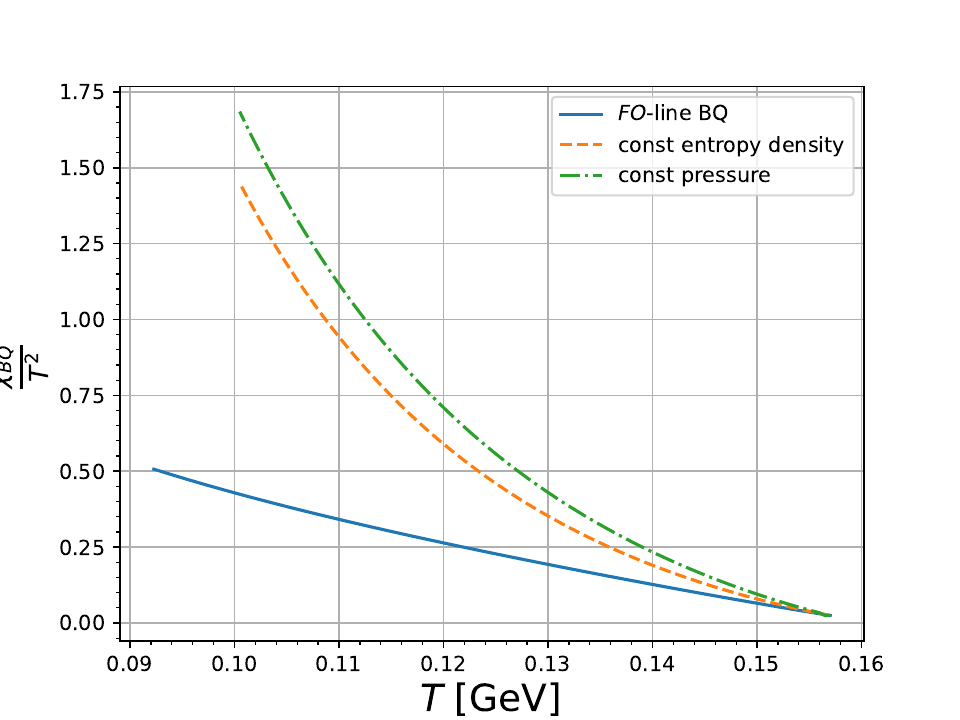}
\includegraphics[width=0.48\textwidth]{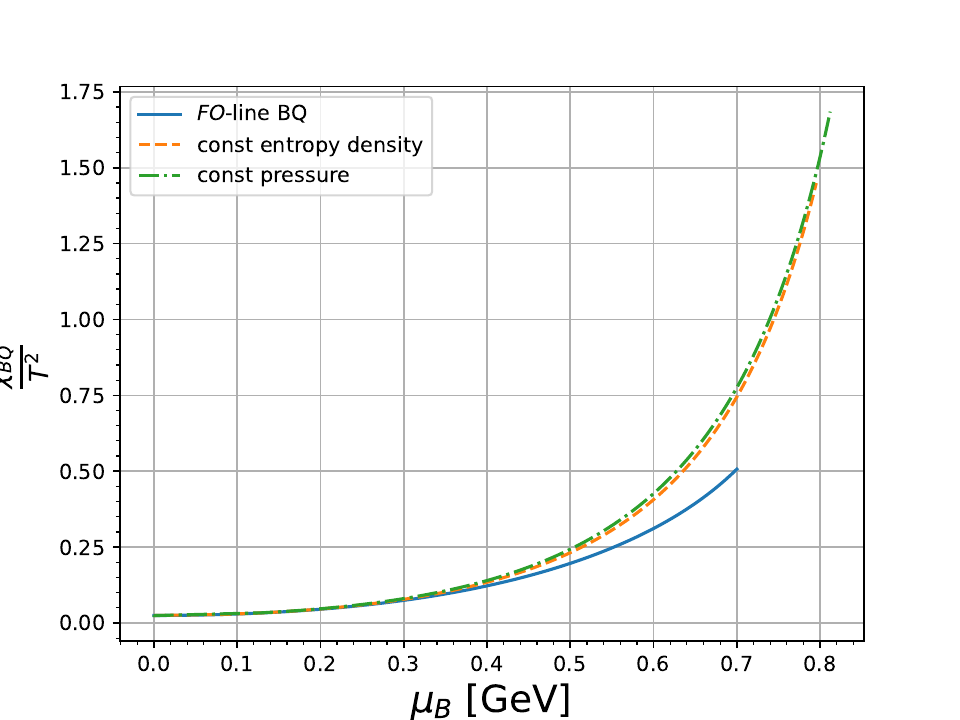}
\end{center}
\caption{$\chi_{BQ}/T^2$ calculated along the FO line, line of constant entropy density, and line of constant pressure. Top: dependence on $T$; bottom: dependence on $\mu_B$.}
\label{f:chiFO}
\end{figure}
%
In Fig.~\ref{f:chiFO} we show $\chi_{BQ}/T^2$ as a function of temperature, calculated along the freeze-out line. 
We compare this with the $T$-dependencies calculated along the lines of constant entropy density and pressure. 
By construction, they start at the same value at $T = 0.157$~GeV but  differ considerably at low temperatures. 
This difference results firstly from the differences of the curves at low $T$ seen in Fig.~\ref{f:fo}, and secondly from the strong dependence of $\chi_{BQ}/T^2$ on the chemical potential seen in Fig.~\ref{f:chiS0}. 
The right panel confirms this by showing the same quantity as function of $\mu_B$: the curves are closer to each other than in the plot of the $T$-dependence.


\section{Partial chemical equilibrium}
\label{s:pce}

While hadron abundances are reproduced by the statistical model with temperature parametrised by eq.~(\ref{e:tfo}) \cite{Stachel:2013zma,Vovchenko:2015idt}, the observed momentum distributions indicate that hadrons are emitted in the final state later, when the temperature has dropped close to 100~MeV \cite{STAR:2008med,ALICE:2013mez,STAR:2017sal,Melo:2019mpn}. 
One possible scenario between the earlier chemical and the later kinetic freeze-out is that of Partial Chemical Equilibrium (PCE) \cite{Bebie:1991ij}. 
After the chemical freeze-out, the effective numbers of stable hadrons---i.e., those measured after the termination of interactions and decay of all resonances---remain constant, even though the temperature drops. 
This leads to the ratios of the effective numbers of different stable hadron species that are not equilibrated at the lower temperature. 
On the other hand, the resonances are in equilibrium with their daughter particles, hence the name Partial Chemical Equilibrium. 
The model is made calculable by the  assumption that the evolution conserves entropy, which introduces some simplification compared to reality. 
Since the entropy density is independently calculated in the model via eq.~(\ref{e:sGD}), this allows to account for the change of the volume. 
Practically, the calculation requires that for each stable hadron species, the ratio
\begin{equation}
\frac{\tilde n_h(T)}{s(T)} = \frac{\tilde n_h(T_{\mathrm{fo}})}{s(T_{\mathrm{fo}})} = \mathrm{const}\, .
\label{e:pcec}
\end{equation}
Here, $\tilde n_h(t)$ include directly produced hadrons, those from the decays of resonances, as well as contribution from the interaction term 
\begin{equation}
\tilde n_h = n_h + \sum_R n_R p_{R\to h} + \sum_{\mathrm{int}} n_h^{\mathrm{int}}\, .
\end{equation}
The second sum goes over all resonance species and $p_{R\to h}$ is the average number of hadrons $h$ produced in decays of resonance $R$.
The third sum goes over all interaction channels accounted for through the phase shifts. 
Note that the splitting into the second and the third term is solely technical, as both these terms represent interactions between hadrons. 

To fulfill the condition (\ref{e:pcec}) while the temperature is decreasing from the chemical freeze-out value, chemical potentials of the individual stable species are evolved. 
Resonances inherit the chemical potentials from the daughter species
\begin{equation}
\mu_R = \sum_h p_{R\to h} \mu_h
\end{equation}
where the sum goes over all stable hadron species to which given resonance decays. 
The phase-shift term in our model only represents  $\pi N$ interactions and the chemical potentials used there are given by the sum of $\mu_\pi$ and $\mu_N$  in the appropriate isospin channel.

The evolution of the chemical potentials based on eq.~(\ref{e:pcec}) is started from temperatures and chemical potentials obtained from fits to hadron abundances by the STAR collaboration \cite{STAR:2017sal}. 
They are listed in Table~\ref{t:tandmu}. 
Strange neutrality has been assumed throughout this calculation. 
\begin{table}[t]
\caption{Initial values of $T$ and $\mu_B$ from which the evolution of PCE chemical potentials is started, for different collision energies. 
Values taken from \cite{STAR:2017sal}.}
\label{t:tandmu}
\begin{tabular}{ccc}
\hline
$\sqrt{s_{NN}}$ [GeV] & $T$ [MeV] & $\mu_B$ [MeV] \\
\hline\hline
7.7	&	144.3	&	398.2\\
11.5	&	149.4	&	287.3\\
19.6	&	153.9	&	187.8\\
27.0	&	155.0	&	144.4\\
39.0	&	156.4	&	103.2\\
62.4	&	160.3	&	69.8\\
200	&	164.3	&	28.4\\
\hline
\end{tabular}
\end{table}
Again, those fits have been performed with the HRG model without explicit phase shifts and with both $\mu_B$ and $\mu_S$ as free parameters, but they are used here  since an analysis with phase shifts is not available and its results are expected to be very close to what we use now. 
%
\begin{figure*}[t]
\begin{center}
\includegraphics[width=0.96\textwidth]{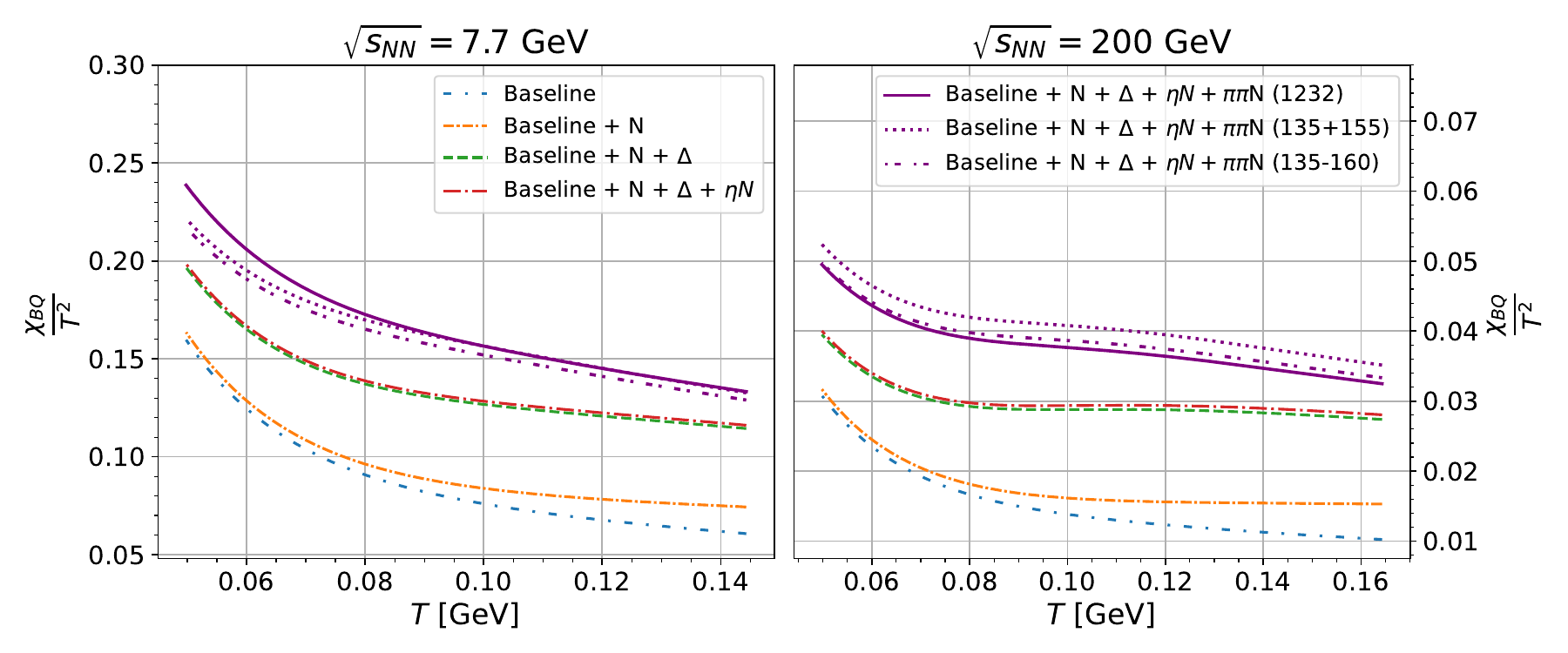}
\end{center}
\caption{Contributions to $\chi_{BQ}/T^2$ calculated as functions of $T$ for cooling within the PCE scenario. 
'Baseline' contribution does not include the phase shifts part of the partition function, 'Baseline + N' adds the $I_z = 1/2$ channels of interactions, 'Baseline + N + $\Delta$' adds the $I=3/2$ channels, and 'Baseline + N + $\Delta$ + $\eta N$' adds the $\eta N$ channel. Purple curves show the results from the three models accounting for $\pi\pi N$ interactions; solid: $\Delta(1232)$ model, dotted: Shifted $N^*$ and $\Delta$ model 1, double-dotted: Shifted $N^*$ and $\Delta$ model 2. Left: cooling from the chemical FO in $\sqrt{s_{NN}}=7.7$~GeV collisions. Right: same for $\sqrt{s_{NN}}=200$~GeV collisions.}
\label{f:pcechicomp}
\end{figure*}
%
In Fig.~\ref{f:pcechicomp} we show $\chi_{BQ}/T^2$ as a function of temperature for two different collision energies from the RHIC BES I programme. 
As the temperature is lowered, chemical potentials were calculated according to the PCE model. 
One can see that in agreement with Fig.~\ref{f:chiS0} the susceptibility is much higher at low collision energy which produces matter at high net baryon density. 
A large contribution to the final result is due to $I=\frac{3}{2}$ interactions, especially at high temperatures. 
Three-body interactions give a comparably large contribution, as well. 
The systematic uncertainty due to different $\pi\pi N$ models accounts to about 20\%. 
This is illustrated by showing results from all three models that were used to describe it. 
The $\chi_{BQ}/T^2$ mostly increases as the temperature is lowered, but when the scaling by $T^2$ is removed, the susceptibility is goes down as the medium cools, as would be expected. 
This is shown in Fig.~\ref{f:pcechinoT2}.
%
\begin{figure*}[t]
\begin{center}
\includegraphics[width=0.96\textwidth]{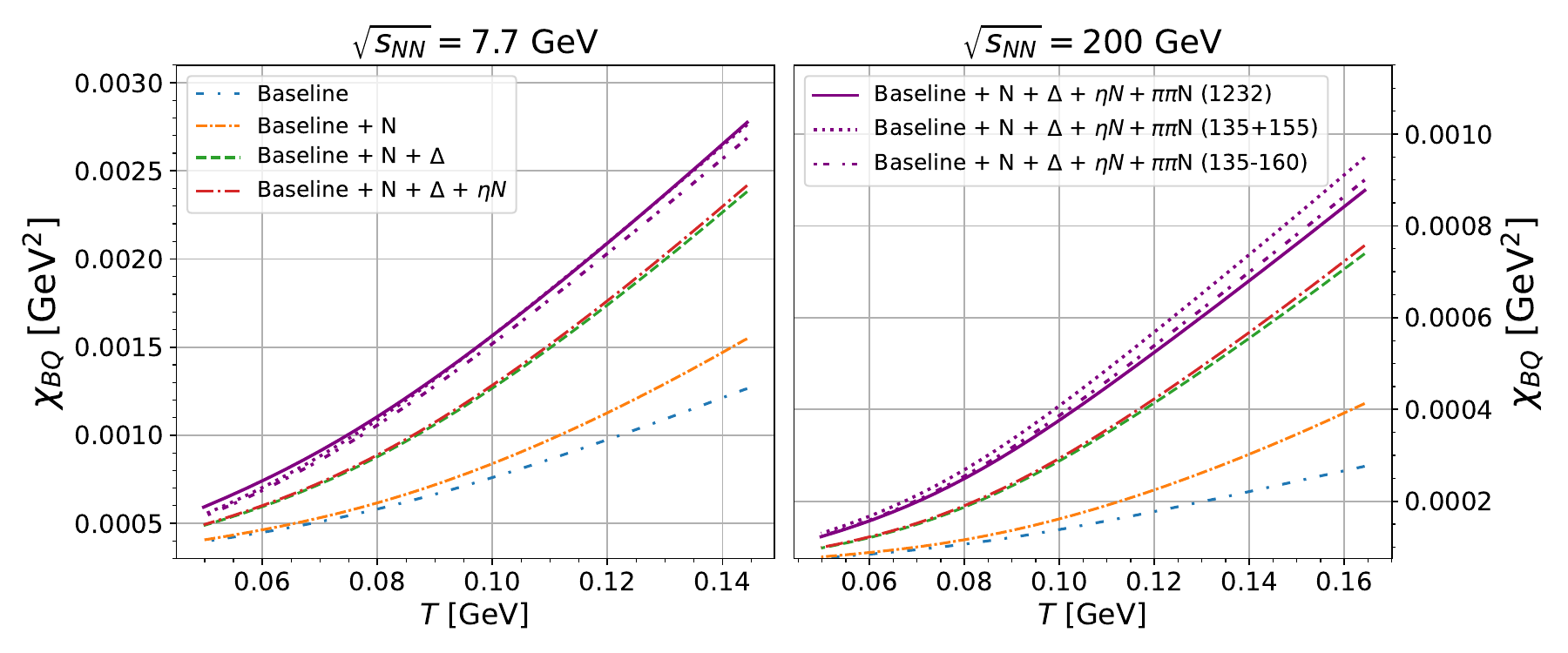}
\end{center}
\caption{Same as Fig.~\ref{f:pcechicomp} but $\chi_{BQ}$ is plotted instead of $\chi_{BQ}/T^2$.}
\label{f:pcechinoT2}
\end{figure*}
%

\begin{figure*}[th]
\begin{center}
\includegraphics[width=0.96\textwidth]{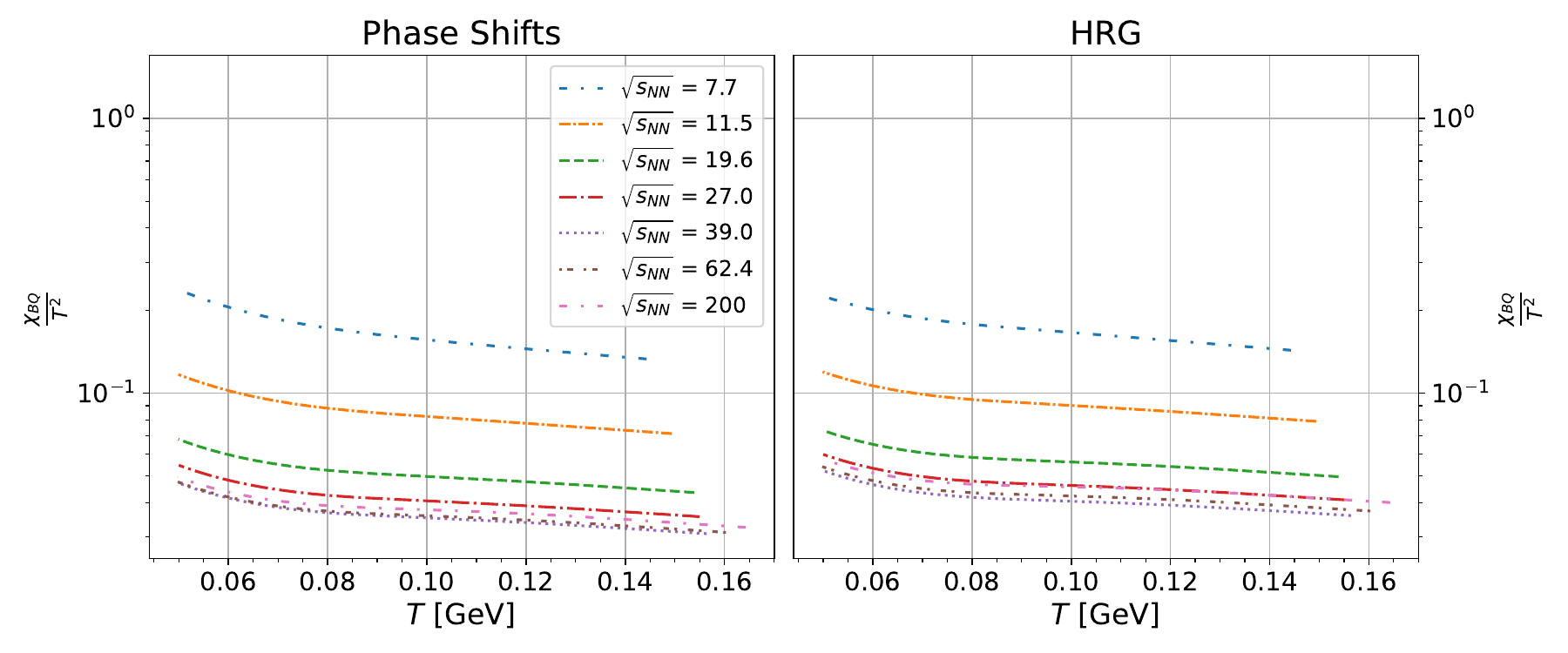}
\end{center}
\caption{Temperature dependence of $\chi_{BQ}/T^2$ for cooling matter within the PCE scenario for chemical FO at various collision energies. Left: calculations with interactions included through phase shifts. Right: Hadron Resonance Gas model with resonances of vanishing widths.}
\label{f:chiall}
\end{figure*}

Susceptibilities for all collision energies are shown in Fig.~\ref{f:chiall}.
%
%
In the logarithmic plot the $T$-dependence of $\chi_{BQ}/T^2$ seems to be universal, but this is only due to limited resolution of such a plot. 
Figure~\ref{f:pcechicomp} clearly shows the difference between the curves at high and low energies. 
%
For comparison, we also show the susceptibilities calculated within the HRG model where resonances with vanishing widths replace the treatment using phase shifts.
The use of phase shifts  lowers the resulting $\chi_{BQ}$.


\section{Conclusions}
\label{s:conc}

We have extended the calculation \cite{Lo:2017lym} of $\chi_{BQ}$ to regimes that are phenomenologically relevant and interesting, because the matter produced in relativistic heavy-ion collisions has non-vanishing net-baryon density. 
Recall that the fluctuations of multiplicities are one of the observables sensitive to the appearance of the critical point and the first-order phase transition at some non-zero baryo-chemical potential. 
From this point of view, what we have calculated would serve as a baseline value and any critical contribution connected with the appearance of the critical point should be on top of this. 

Our results show that at higher baryon density, accessible in collisions at lower energies, the $BQ$ susceptibility grows considerably. 
The $S$-matrix treatment of the $\pi N$ interaction makes the result slightly smaller than the calculation with simple HRG model, but the difference is not dramatic comparing to the large value of $\chi_{BQ}/T^2$. 

Nevertheless, the phenomenologically relevant benchmark value of $\chi_{BQ}$ may be lower than what is calculated at the chemical freeze-out. 
Our calculation using the PCE model indicates that the cooling in hadronic phase may lead to considerable decrease of $\chi_{BQ}$. 
For example, it drops to about 60\% of its chemical FO value when temperature is lowered to 100~MeV, as found for $\sqrt{s_{NN}}=7.7$~GeV (Fig.~\ref{f:pcechinoT2}).
This should be taken into account when the measured $BQ$ correlations are being interpreted. 


\backmatter

\bmhead{Acknowledgement}
BT acknowledges  support from the Czech Science Foundation under grant No 25-16877S. 
PML acknowledges partial support from the Polish National Science Center (NCN) under the Opus grant no. 2022/45/B/ST2/01527.


\end{document}